\documentclass[aps,prl,10pt,superscriptaddress,twocolumn,footinbib]{revtex4-2}
\usepackage[cm]{fullpage}
\usepackage{microtype}
\usepackage{amsfonts,amsmath,amssymb,bm,mathtools,physics}
\usepackage{graphicx,xcolor}
\usepackage[pdfstartview=FitH,colorlinks=true,citecolor=blue,linkcolor=blue,urlcolor=blue]{hyperref}

\begin{document}

\newcommand{\rperp}{\mathbf{r}_\perp}
\newcommand{\kperp}{\mathbf{k}_\perp}

\title{
Swimming against a superfluid flow: \texorpdfstring{\\}{Lg} Self-propulsion via vortex-antivortex shedding in a quantum fluid of light
}

\author{Myrann Baker-Rasooli}
\affiliation{Sorbonne Universit\'e, ENS Paris, Coll\`ege de France, Universit\'e PSL, CNRS, Laboratoire Kastler Brossel, F-75005 Paris, France}

\author{Tangui Aladjidi}
\affiliation{Sorbonne Universit\'e, ENS Paris, Coll\`ege de France, Universit\'e PSL, CNRS, Laboratoire Kastler Brossel, F-75005 Paris, France}

\author{Tiago D. Ferreira}
\affiliation{Universidade do Porto and INESC TEC, Centre of Applied Photonics, Rua do Campo Alegre, Porto, Portugal}

\author{Alberto Bramati}
\affiliation{Sorbonne Universit\'e, ENS Paris, Coll\`ege de France, Universit\'e PSL, CNRS, Laboratoire Kastler Brossel, F-75005 Paris, France}
\affiliation{Institut Universitaire de France (IUF)}

\author{Mathias Albert}
\affiliation{Universit\'e C\^ote d'Azur, CNRS, INPHYNI, France}
\affiliation{Institut Universitaire de France (IUF)}

\author{Pierre-\'Elie Larr\'e}
\email{pierre-elie.larre@universite-paris-saclay.fr}
\affiliation{Universit\'e Paris-Saclay, CNRS, LPTMS, 91405, Orsay, France}

\author{Quentin Glorieux}
\email{quentin.glorieux@sorbonne-universite.fr}
\affiliation{Sorbonne Universit\'e, ENS Paris, Coll\`ege de France, Universit\'e PSL, CNRS, Laboratoire Kastler Brossel, F-75005 Paris, France}
\affiliation{Institut Universitaire de France (IUF)}

\date{\today}

\begin{abstract}
A superfluid flows without friction below a critical velocity, exhibiting zero drag force on impurities. Above this threshold, superfluidity breaks down, and the internal energy is redistributed into incoherent excitations such as vortices. We demonstrate that a  finite-mass, mobile impurity immersed in a flowing two-dimensional paraxial superfluid of light can \textit{swim} against the superfluid current when this critical velocity is exceeded. This self-propulsion is achieved by the periodic emission of quantized vortex-antivortex pairs downstream, which impart an upstream recoil momentum that results in a net propulsive force. Analogous to biological systems that minimize effort by exploiting wake turbulence, the impurity harnesses this vortex backreaction as a passive mechanism of locomotion. Reducing the impurity dynamics to the motion of its center of mass and using a point-vortex model, we quantitatively describe how this mechanism depends on the impurity geometry and the surrounding flow velocity. Our findings establish a fundamental link between internal-energy dissipation in quantum fluids and concepts of self-propulsion in active-matter systems, and opens new possibilities for exploiting vortices for controlled quantum transport at the microscale.
\end{abstract}

\maketitle

In normal fluids, viscous drag opposes the motion of immersed impurities, so efficient motion requires mitigating energy losses due to these friction forces. In nature, a wide variety of strategies have evolved to address this constraint: bacteria and microalgae rely on chemotaxis to navigate chemical gradients~\cite{brennen1977fluid, xie2011bacterial, bureau2023lift}, birds and fishes extract energy from wavy or turbulent streams~\cite{wu1975extraction, fausch1984profitable, drucker2001locomotor}, and bioinspired microrobots exploit similar hydrodynamic phenomena~\cite{palagi2018bioinspired}. Topology also enables efficient transport in active matter~\cite{bowick2022symmetry}, and even in antiferromagnets through self-propulsion of skyrmion textures~\cite{de2025emergent}. In an even more striking illustration, a dead trout has been shown to passively drift upstream~\cite{liao2003fish, liao2006} by harvesting momentum from K\'arm\'an vortex streets shed in a river flow~\cite{vonkarman1963aerodynamics, cooper2001aeroelastic}.

How do these concepts translate to quantum fluids and, in particular, can one swim in a superfluid? A superfluid has the fundamental property that below a critical velocity, it can flow without friction, suppressing drag on obstacles. First discovered in liquid helium~\cite{leggett1999superfluidity, balibar2007discovery}, superfluidity has since been observed in ultracold atomic gases~\cite{PhysRevLett.83.2502} and, more recently, nonlinear optical media~\cite{amo2009superfluidity, michel2018superfluid}. Above the critical velocity, superfluidity breaks down, marking the onset of dissipation. In two dimensions (2D), this phenomenon typically manifests by the nucleation of quantized vortex-antivortex pairs in the wake of a wide, impenetrable impurity~\cite{Frisch1992}, at velocities lower than Landau's speed of sound~\cite{PhysRevA.91.053615, PhysRevA.107.023310, Eloy2021, Huynh2024}. While this dissipative effect is well understood for fixed, infinite-mass impurities, the case of a mobile, finite-mass impurity~\cite{Luis2020, Misha2021, Misha2024} remains largely unexplored, presenting an open question regarding its dynamic behavior in this specific 2D supercritical regime.

Over the past decade, paraxial superfluids of light~\cite{glorieux2025paraxial} have emerged as a main platform for exploring complex quantum hydrodynamics in 2D~\cite{Larre2015, Vocke2016, fontaine2018observation, Rodrigues2020, Situ2020}. In these systems, light propagation within a Kerr medium is governed by the 2D nonlinear Schr\"odinger (NLS) equation~\cite{Boyd2020}, an analogue of the Gross-Pitaevskii (GP) equation for dilute Bose-Einstein condensates~\cite{pitaevskii2016bose}. This equivalence permits the observation of fundamental 2D superfluid phenomena with classical light, encompassing zero drag~\cite{michel2018superfluid}, vortex turbulence~\cite{Eloy2021, baker2023turbulent}, and Jones-Roberts solitons~\cite{baker2025observation}. 

In this work, we use the paraxial superfluid of light platform, specifically realized in a hot rubidium vapor~\cite{glorieux2023hot}, to experimentally investigate the fundamental problem of a mobile impurity immersed in a 2D superfluid.
We develop an new optical setup to monitor the full dynamics of an all-optical impurity by simultaneously recording its trajectory along the hot atomic medium, and measuring both the amplitude and phase of the superfluid of light at the exit of the medium.
By submitting the impurity to a transverse superfluid flow, we observe a counter-intuitive effect above the superfluid critical velocity: the impurity moves upstream, in stark contrast to the downstream displacement reported in a similar experiment~\cite{michel2018superfluid}. 
Phase-resolved measurements reveal that this upstream motion coincides with the periodic nucleation of vortex-antivortex pairs downstream of the impurity, triggered when the injected flow velocity exceeds the critical speed. 
We show that the recoil produced by each emission event generates a counterflow that helps to propel the impurity against the superfluid current, providing a quantum-hydrodynamics analogue of passive vortex-powered swimming in classical fluids. 
Our work investigates and quantifies, both experimentally and theoretically, this vortex-induced counterflow mechanism, depending critically on the impurity's radius and on the surrounding flow velocity, contrasting it with the well-established sonic drag observed in other regimes~\cite{michel2018superfluid}.
These findings establish an interesting connection between superfluid hydrodynamics and active-matter propulsion mechanisms, but also vortex-induced motions of bodies in hydrodynamic engineering, highlighting a fundamental, domain-independent role for vortices as momentum-transfer agents.

The propagation of a paraxial, continuous-wave laser beam through a self-defocusing local Kerr medium is governed by the 2D NLS equation for the slowly varying envelope $E(\mathbf{r}_\perp,z)$ of its complex-valued electric field:
\begin{equation}
i \partial_z E = \bigg[{-}\frac{\nabla_{\perp}^{2}}{2 k_0} + V(\mathbf{r}_\perp,z) - k_0 n_2 |E|^2 \bigg] E.
\label{eq:NLSE}
\end{equation}
In this equation, $0<z<L$ is the propagation coordinate along the medium and $\mathbf{r}_\perp=(x,y)$ are the transverse coordinates. The in-medium propagation wavenumber is $k_{0}=2\pi n_{0}/\lambda$, where $n_{0}$ and $\lambda$ are the mean refractive index and the wavelength in free space, and $n_2<0$ is the Kerr nonlinear coefficient of the vapor, here normalized by $n_0$. By analogy with the GP equation, Eq.~\eqref{eq:NLSE} can be interpreted as governing the evolution, in the effective time variable $z$, of the wave function $E=\sqrt{\rho}\exp(i\phi)$ of a 2D superfluid of weakly interacting photons of mass $k_0$. Here, $\rho$ and $\mathbf{v}=\nabla_{\perp}\phi/k_0$ represent the local and instantaneous density and velocity, respectively, of this 2D superfluid of light. In this formal analogy, the speed of sound in the superfluid is defined as $c_\mathrm{s}=(|n_2|\rho_{0})^{1/2}$, where $\rho_0$ is the unperturbed density at the center of the transverse light spot~\cite{fontaine2018observation,piekarski2021measurement}. 
We also introduce the transverse characteristic length $\xi=1/(2k_{0}^{2}|n_{2}|\rho_{0}^{\vphantom{2}})^{1/2}=1/(\sqrt{2}k_{0}c_{\mathrm{s}})$, known as the healing length, and the longitudinal characteristic length $z_{\mathrm{NL}}=1/(k_{0}|n_{2}|\rho_{0})=1/(k_{0}c_{\mathrm{s}}^{2})$, known as the nonlinear length and whose inverse is a similar to a chemical potential for the fluid of light \cite{glorieux2025paraxial}.

Crucially, $V(\mathbf{r}_\perp,z)$ models a repulsive impurity in the superfluid, which consists of a localized refractive index depletion in the transverse $x{-}y$ plane. This impurity can be mobile (i.e., of finite mass) or fixed (infinite mass) with controlled strength and width, thus allowing for the study of the impurity problem over a broad range of parameters. In optics, a localized change of the refractive index can be induced by an external field, that we will denote the impurity beam, via cross-phase modulation (XPM)~\cite{Boyd2020}. 
In full generality, when two fields are coupled via XPM, the system is described by two coupled NLS equations. 
However, if we assume that the impurity beam maintains its transverse shape for all $z$, we can approximate the potential $V$ by $V(\mathbf{r}_\perp, z) = g_{\mathrm{XPM}} |E_{\mathrm{i}}|^2$, with $E_{\mathrm{i}}(\mathbf{r}_\perp, z)$ the (complex-valued) electric-field envelope of the impurity beam, and $g_{\mathrm{XPM}}$ the XPM coupling. 
This approximation is typically realized in nonlinear media when the paraxial-diffraction, kinetic-energy term is exactly compensated by the self-focusing, attractive-interaction term, forming a bright-solitonlike beam.
In this case, the coupled NLS equations simplify and the dynamics of the impurity beam is given by a force deriving from a potential coming from the XPM induced by the fluid-density profile:
\begin{equation}
\label{Eq:XPM}
i \partial_z E_{\mathrm{i}} = g_{\mathrm{XPM}} |E|^2 E_{\mathrm{i}}.
\end{equation}

\begin{figure*}[t!]
    \centering
    \includegraphics[width=\linewidth]{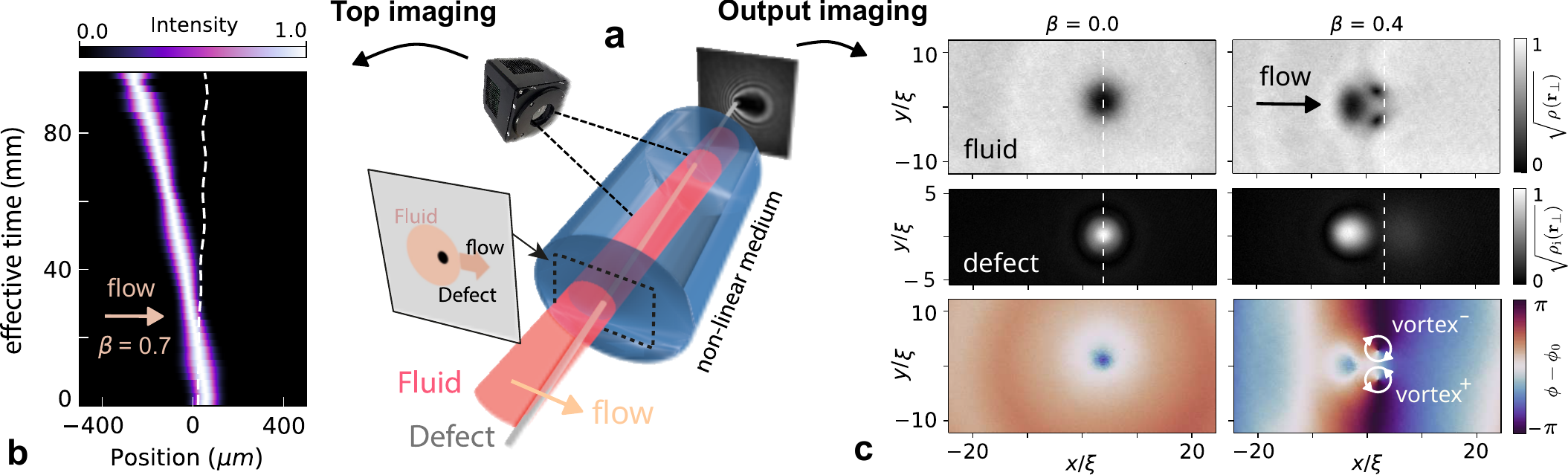}
    \caption{\textbf{Hydrodynamic flow of superfluid light past a mobile optical impurity.}
    \textbf{a} -- Schematic of the experimental setup.
    The fluid beam is generated with a 780 nm laser beam, sent through a 20 cm-long hot rubidium-vapor cell maintained at 150 °C.
    The optical impurity is generated by injecting a narrower, 795 nm beam, which overlaps the fluid beam.
    The relative angle between the two beams controls the transverse flow velocity.
    \textbf{b} -- Intensity of the impurity beam, measured along the rubidium cell using the top camera and a frequency filter at 795 nm.
    Each segment is normalized relative to its maximum value.
    Here, the Mach number of the flow is $\beta = 0.7$. 
    The observed shift of the beam relative to its smoothed white dashed reference position, corresponding to the fluid of light at rest, indicates that the impurity moves upstream in the transverse plane.
    \textbf{c} -- Transverse intensity of the fluid of light at the cell output for $\beta = 0$ and $\beta = 0.4$.
    Top: fluid amplitudes taken with a 780 nm filter.
    Middle: impurity amplitudes taken with a 795 nm filter.
    Bottom: associated phase of the fluid. 
    Each amplitude image is normalized to its maximum amplitude value.
    The vortex-antivortex pair generated downstream at $\beta=0.4$ is highlighted with white circles centered on the $\pm2\pi$ phase-windings, respectively.}
    \label{fig:setup_swim}
\end{figure*}

In this work, we implement experimentally this superfluid model in the case of an impurity of finite mass $k_{\mathrm{i}}$. This is realized by propagating two XPM-coupled laser beams through a $L=20$ cm-long, \textsuperscript{87}Rb-vapor cell heated to $150~^{\circ}\mathrm{C}$~\cite{glorieux2023hot}, as illustrated in Fig.~\ref{fig:setup_swim}(a).
The fluid beam described by Eq.~\eqref{eq:NLSE} corresponds to a $\lambda=780~\mathrm{nm}$ laser beam, slightly detuned from the rubidium D2 line ($\sim-8$~GHz). At this wavelength, the warm rubidium vapor displays a substantial Kerr nonlinearity which induces repulsive photon interactions and therefore creates a dynamically stable superfluid~\cite{fontaine2018observation}.
The impurity beam described by Eq.~\eqref{Eq:XPM}, and which produces an inhomogeneity $V(\mathbf{r}_{\perp},z)$ for the fluid evolution in~\eqref{eq:NLSE}, operates at $\lambda_{\mathrm{i}}=2\pi n_{0}/k_{\mathrm{i}}=795~\mathrm{nm}$, tuned close to the D1 line with a positive detuning ($\sim+2$~GHz) to maintain its shape during propagation.
Since both beams share the same ground state, the impurity will induce optical pumping of the atoms (locally) and therefore induces a XPM on the fluid beam.
The impurity has a radius (laser waist) $\sigma\simeq 7.5\xi$ of the order of a few healing lengths $\xi \simeq 50~\mu\mathrm{m}$, much smaller than the overall superfluid size in the transverse plane, $\simeq 100\xi$.
By optically adjusting the relative angle between the fluid and impurity beams, we adjust, by definition of the local flow velocity $\mathbf{v}$, the incoming flow velocity $v_{0}$, here in the positive-$x$ direction. Measured relatively to the speed of sound $c_{\mathrm{s}}$, it defines the Mach number of the flow, $\beta=v_{0}/c_{\mathrm{s}}$ [see Supplemental Material (SM) for details].

Using a dual-imaging system, we simultaneously measure the fluorescence of the fluid beam from the top and record the amplitude $\sqrt{\rho}$ and phase $\phi$ of the superfluid of light at the exit of the rubidium cell. 
As illustrated in Fig.~\ref{fig:setup_swim}(b), we track the impurity trajectory with an injected initial flow ($\beta=0.7$ on the figure), and observe a clear deviation compared to the situation where the superfluid is at rest ($\beta=0$), indicated by the smoothed white dashed line (its wiggly nature is due to the inconsistencies in the imaging train positioning).
The output amplitude and phase of the fluid beam are recorded using an off-axis interferometer. 
Typical images are shown in Fig.~\ref{fig:setup_swim}(c), where we display the fluid amplitude $\sqrt{\rho}$ (top panels) for $\beta=0$ and $\beta=0.4$, together with the corresponding phase maps $\phi$ relative to the phase $\phi_{0}$ in the absence of the impurity (bottom). 
For $\beta=0.4$, a vortex-antivortex pair is observed in the wake of the impurity, identified by detecting $ \pm 2\pi $ phase windings.
Importantly in the present study, direct access to the fluid phase allows for a measurement of the local flow velocity, $\mathbf{v}(\mathbf{r}_{\perp})$ (see SM for details), thus making it possible to compute the vorticity $\nabla_\perp \times \mathbf{v}(\mathbf{r}_\perp)$ of the superfluid.
Additionally, by applying a narrow frequency filter at 795 nm, we extract the amplitude profile of the impurity, $\sqrt{\rho_{\mathrm{i}}}=|E_{\mathrm{i}}|$, at the exit of the cell, as shown in Fig.~\ref{fig:setup_swim}(c) (middle panels), and revealing its displacement in the transverse plane as $\beta$ increases.

\begin{figure*}[t!]
    \centering
    \includegraphics[width=\linewidth]{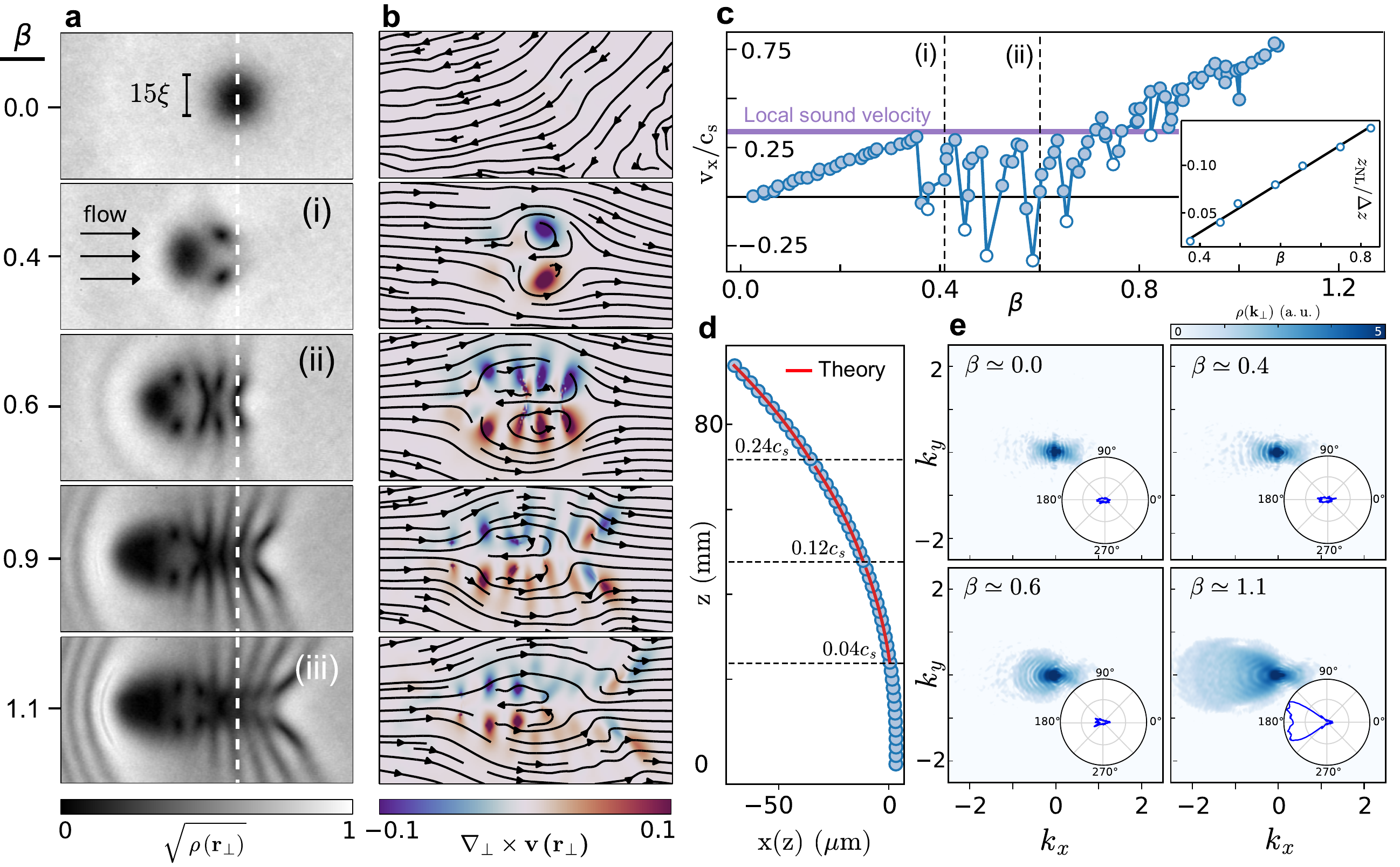}
    \caption{\textbf{Upstream motion of the impurity by vortex-antivortex shedding.}
    \textbf{a} -- Output amplitude images of the fluid for increasing values of the incoming Mach number, from $\beta=0$ (top) to $\beta=1.1$ (bottom), with each image normalized by its maximum value.
    The typical radius of the impurity is $\simeq7.5\xi$.
    The dashed line indicates the initial position of the impurity, highlighting upstream motion from vortex shedding.
    \textbf{b} -- Associated vorticity and streamlines maps.
    A counterflow appears along the central axis of the vortex alley produced downstream.
    \textbf{c} -- Local flow velocity along the $x$ axis at the position of the impurity as a function of the incoming Mach number $\beta$.
    When the local flow velocity reaches the local sound velocity ($\simeq0.35c_{\mathrm{s}}$; thick purple line)---which occurs at $\beta\equiv\beta_{\mathrm{c}}\simeq0.35$---the flow transitions from superfluid to normal. This change is marked by the periodic emission of vortex-antivortex pairs (white circles) and the onset of a counterflow at the center of these pairs.
    The two vertical dashed lines correspond to images~(i) and~(ii) on panel~(a), respectively.
    The inset shows the vortex-shedding frequency $1/\Delta z$ against $\beta$, obtained at each emission event. The solid line is obtained from a linear fit based on prediction~\eqref{Eq:Deltaz}.
    \textbf{d} -- Impurity trajectory of Fig.~\ref{fig:setup_swim}(b) measured along the rubidium cell at $\beta=0.7$.
    The red lines show the fits obtained using Eqs.~\eqref{Eq:Xz} and~\eqref{Eq:DensityImbaba} at each vortex emission.
    \textbf{e} – Momentum distribution of the fluid's excitations in the upstream region in the reference frame of the fluid for $\beta=0,~0.4,~0.6,~1.1$.
    The inset of each image shows the associated polar plot.}
    \label{fig:vortex_swim}
\end{figure*}

Above a critical velocity, we observe that the impurity sheds a trailing wake of vortex-antivortex pairs orthogonal to the incident flow, as quite expected in 2D superfluids, while strikingly moving upstream, \textit{opposite} to the incoming stream. This is illustrated in Fig.~\ref{fig:vortex_swim}.
Panel~(a) shows amplitude images of the superfluid of light past the impurity for Mach numbers varying from $\beta=0$ to $\beta=1.1$ (top to bottom).
The corresponding vorticity maps are displayed in panel~(b), with the associated streamlines.
The colormap (saturated for visibility) highlights the downstream nucleation of paired vortices (in red; $+1$ winding number) and antivortices (blue; $-1$), while revealing the formation of a counterflow, in the negative-$x$ direction, along the central axis of the vortex alley formed downstream.
In Fig.~\ref{fig:vortex_swim}(c), we measure the $x$-component of the local superfluid velocity in $c_{\mathrm{s}}$ units at the maximum of the impurity potential, $v_{x}/c_{\mathrm{s}}$, as a function of the injected Mach number $\beta$.
The horizontal solid line in black indicates $0$ and the two vertical dashed lines correspond to images~(i) ($\beta=0.4$) and~(ii) ($\beta=0.6$) in panel~(a).
As expected theoretically for a 2D potential flow around a circular cylinder~\cite{landau1987fluid}, the local flow velocity $v_{x}$ increases linearly with $\beta$ until reaching the \textit{local} speed of sound $\tilde{c}_{\mathrm{s}}(\mathbf{r}_{\perp})=c_{\mathrm{s}}[\rho(\mathbf{r}_{\perp})/\rho_{0}]^{1/2}$ at the center of the impurity, indicated by the horizontal solid line in purple. 
At the corresponding Mach number, $\beta\equiv\beta_{\mathrm{c}}$, a pair of vortex-antivortex is emitted downstream, marking the transition from super- to dissipative flow. 
This is consistent with the \textit{local} Landau criterion, which predicts, for wide impurities ($\sigma\gg\xi$; $\sigma/\xi\simeq7.5$ in our experiment), that superfluidity breaks down when the local flow velocity reaches the local sound speed at the impurity position~\cite{Pomeau1993, Hakim1997}.
The measured critical Mach number for vortex nucleation, $\beta_{\mathrm{c}} \simeq 0.35$, is consistent with theoretical results for a wide impurity (see recent work~\cite{Huynh2024} and references therein). A comprehensive study detailing the dependence of our $\beta_{\mathrm{c}}$ on the impurity parameters is the subject of a companion work~\cite{TanguiCrit}.

At each vortex-antivortex emission, indicated by white circles on panel~(c), the counterflow observed in panel~(b) locally reduces the velocity below the local sound velocity.
Assuming that the vortex pairs are emitted periodically~\cite{lim2022vortex}, we can extract the vortex-shedding frequency $1/\Delta z$ (see SM for details). 
We find this frequency to linearly increase with the incoming Mach number, following
\begin{equation}
\label{Eq:Deltaz}
\frac{1}{\Delta z}=\frac{\sqrt{2}a\xi}{z_{\mathrm{NL}}}(\beta-\beta_{\mathrm{c}}),
\end{equation}
where $a$ is some inverse-length constant.
The inset of Fig.~\ref{fig:vortex_swim}(c) shows this frequency in units of the inverse nonlinear length, measured as a function of the $\beta$'s marking the nucleation of the vortex pairs. It exhibits a linear trend (solid line; $a\xi \simeq 0.19$) that is consistent with cold-atom experiments~\cite{kwon2015periodic, lim2022vortex}.

The observed upstream motion of the impurity, depicted for instance in the trajectory $X=X(z)$ of Fig.~\ref{fig:setup_swim}(b) for $\beta=0.7$, is manifestly associated with the periodic emission of vortex-antivortex pairs in the wake of the impurity. A theoretical model is crucial for providing analytical insight into the physical origin of this phenomenon. We test the hypothesis that the emitted vortex-antivortex pairs accelerate the impurity upstream by modeling its trajectory using a phenomenological description of the coupled superfluid-impurity dynamics right above the critical speed for vortex nucleation. This upstream acceleration is fundamentally related to the drag force experienced by the impurity. As detailed in the SM, this force is estimated within a simplified model that reduces the impurity dynamics to the motion of its center of mass, coupled to the NLS equation describing the superfluid. Approximations include the use of the point-vortex model~\cite{Dalibard2017, Skipp2023} to calculate the velocity field generated by the vortices emitted downstream. Within this framework, the impurity's trajectory after each vortex emission is approximated by a parabola in the effective time $z$ (plus linear and constant terms):
\begin{equation}
\label{Eq:Xz}
X(z)=Az^{2}+\dot{X}_{0}z+X_{0},\quad A=-\frac{\alpha\ell V_{0}\Delta\rho}{k_{\mathrm{i}}}.
\end{equation}
Here, $X_{0}$ and $\dot{X}_{0}$ are the position and velocity of the impurity at the start of a vortex emission. The acceleration $A$ depends on several key quantities: a constant $\alpha$ which properly normalizes the involved Hamiltonians; the typical transverse size of the incoming flux, $\ell$, the maximum amplitude of the impurity's potential, $V_{0}$, and the propagation constant of the impurity beam, $k_{\mathrm{i}}$. Critically, the sign of the acceleration $A$ in the second of Eqs.~\eqref{Eq:Xz} is determined by the density imbalance $\Delta\rho=\rho_{\mathrm{down}}-\rho_{\mathrm{up}}$ between the downstream ($\rho_{\mathrm{down}}$) and upstream ($\rho_{\mathrm{up}}$) regions of the impurity along the incoming stream of density $\rho_{0}$, which is demonstrated to be given by
\begin{equation}
\label{Eq:DensityImbaba}
\Delta\rho=2\sqrt{2}\rho_{0}\frac{\xi}{\sigma}\bigg(\beta-\sqrt{2}\frac{\xi}{\sigma}\bigg).
\end{equation}
The acceleration $A$ is negative---signifying propulsion against the superflow---provided the term in parentheses is positive, which implies the following condition:
\begin{equation}
\label{Eq:Counterflow}
\beta>\sqrt{2}\frac{\xi}{\sigma}.
\end{equation}
In our experiment, we operate at $\sqrt{2}\xi/\sigma\simeq0.19$ ($\sigma\simeq7.5\xi$), which is below the critical Mach number $\beta_{\mathrm{c}}\simeq0.35$ for vortex nucleation. Condition~\eqref{Eq:Counterflow} is thus satisfied ($\beta\gtrsim\beta_{\mathrm{c}}$), quantitatively confirming that our system is operating in the counterflow-propulsion regime. This self-propulsion is intrinsically due to the formation of a \textit{density hump downstream} ($\Delta\rho>0$), most probably resulting from the upper-clockwise and lower-anticlockwise vorticity windings that effectively draws light intensity towards the rear of the impurity. This density accumulation acts as a sort of localized pressure gradient that pushes the impurity upstream. To validate the model, we employ the previous assumption of periodic emission. The rubidium cell is discretized into equal segments indicating the emission times of the vortices observed at a given $\beta$ [see Fig.~\ref{fig:vortex_swim}(d) for $\beta=0.7$, and details in the SM]. In each segment, Eqs.~\eqref{Eq:Xz} and~\eqref{Eq:DensityImbaba} are adjusted to the measured trajectory, treating the normalization constant $\alpha$ as a fitting parameter (with $X_{0}$ and $\dot{X}_{0}$ obtained from the initial conditions in each emission event). The fit yields $\alpha\simeq3~\mathrm{V}^{-2}$, which is of a nonextreme order of magnitude. This is compelling evidence that all other dependencies within the expression for $A$ accurately account for the dominant physics governing the phenomenon. We note that $\Delta\rho$ vanishes as the impurity's radius $\sigma$ increases. This occurs because a larger diameter forces the vortices in each emitted pair to separate further. This separation reduces the countercurrent generated along the central axis of the vortex alley, thereby reestablishing a density balance ($\Delta\rho\to0$) around the impurity.

Our analysis neglects other sources of drag, such as those arising from sound waves, which typically appear upstream and tend to propel the mobile impurity downstream~\cite{michel2018superfluid} (see SM for discussion). This simplification is supported by Fig.~\ref{fig:vortex_swim}(e), which shows the momentum distributions of the fluid's excitations at different $\beta$'s in the usptream region (in the reference frame of the fluid). 
For $\beta<0.8$, in the vortex-emission regime, there is no significant signature of sound radiation. 
Evidently, sound emission begins at larger velocities, typically when $v_{0}\gtrsim c_{\mathrm{s}}$. 
This onset is signaled by a clear increase in the negative-$k_{x}$ components of the momentum distributions when $\beta$ is of the order of 1, and is also visible in the amplitude images of Fig.~\ref{fig:vortex_swim}(a) as periodic undulations ahead of the impurity for $\beta=0.9$ and $\beta=1.1$.

\begin{figure}[t!]
\centering
\includegraphics[width=\linewidth]{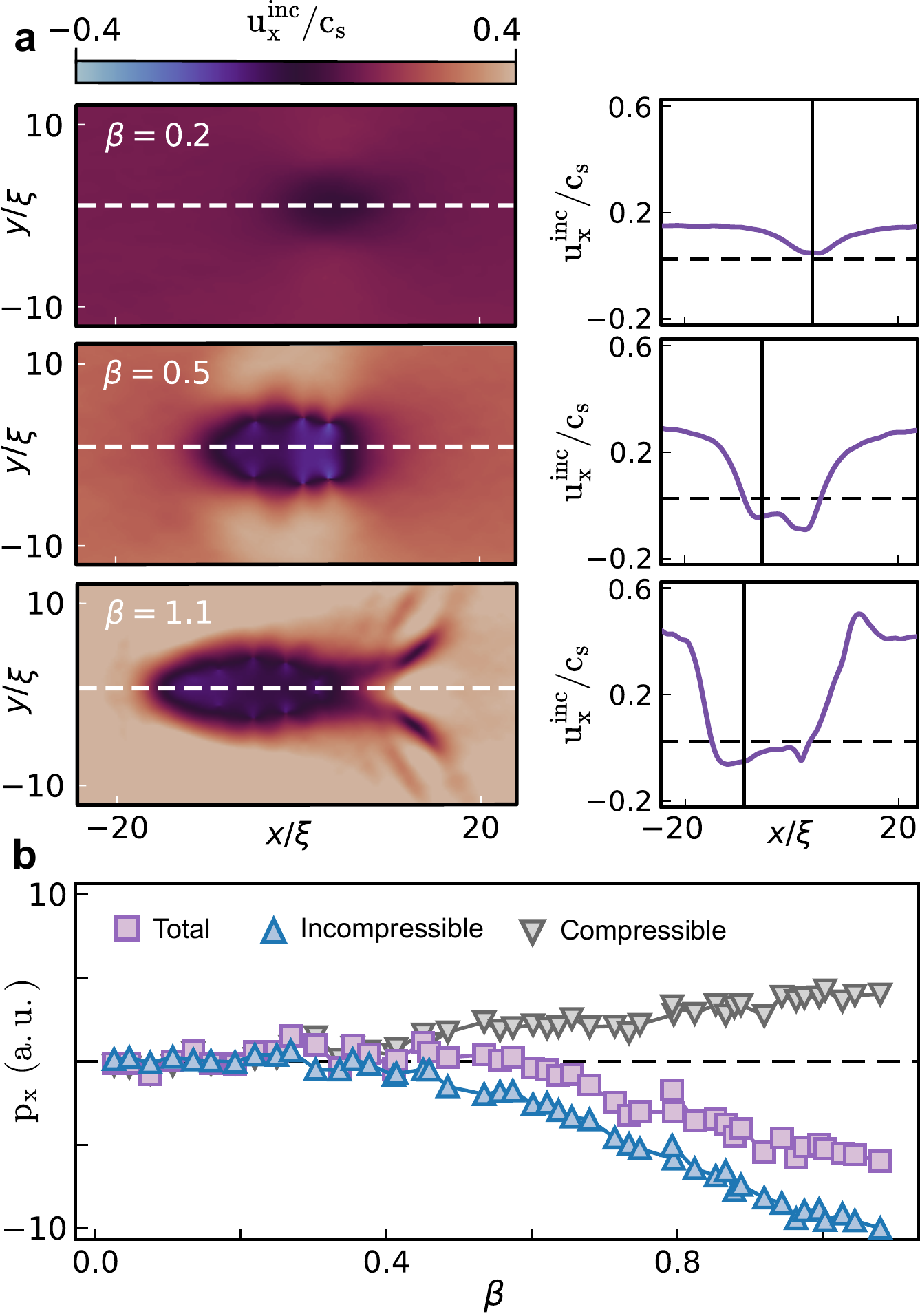}
\caption{\textbf{Density-weighted velocity and net momentum.}
\textbf{a} -- $x$ component of the incompressible density-weighted velocity for $\beta=0.2$ (top), 0.5 (middle), and 1.1 (bottom). The respective right panels show cuts along the white dashed lines at $y=0$. The dark solid lines indicate the impurity position and the dark dashed lines correspond to $u_{x}^{\mathrm{inc}}(x,0)=0$.
\textbf{b} -- Downstream $x$ component of the net momentum of the fluid as a function of $\beta$, in the reference frame of the fluid. Purple squares, blue triangles and gray reversed triangles respectively show the total, incompressible (vortex), and compressible (sound) contributions.}
\label{fig:momentum_swim}
\end{figure}

We conclude this study by providing further quantitative support for the mechanisms elucidated above. We begin by analyzing the density-weighted velocity, defined in terms of the physical velocity $\mathbf{v}$ as $\mathbf{u}=\sqrt{\rho}\mathbf{v}$. The compressible component of $\mathbf{u}$, associated with sound waves, is subtracted to isolate the incompressible field $\mathbf{u}^{\mathrm{inc}}$, which is associated with vortices~\cite{Lamb1932} (see SM for details). The left panels in Fig.~\ref{fig:momentum_swim}(a) display maps of its $x$ component normalized by the speed of sound, $u_{x}^{\mathrm{inc}}(\mathbf{r}_{\perp})/c_{\mathrm{s}}$, at the exit of the cell for $\beta=0.2$, 0.5, and 1.1 (from top to bottom). The colormaps provide distinction between regions where the flow moves in the positive-$x$ direction (hot colors) and those where a counterflow emerges (cold colors). The corresponding line profiles along the white dashed cuts at $y=0$ are shown in the right panels. In the downstream region where vortices are nucleated, these profiles reveal pronounced negative $u_{x}^{\mathrm{inc}}$'s, consistent with the dark-purple regions in the colormaps, thus confirming the presence of a local negative flow induced by vortex dynamics, along the central, $y=0$ axis of the vortex alley.

We also analyze the system's net momentum in the downstream region, which we define from $\mathbf{u}$ as $\mathbf{p}=\int_{\mathrm{down}}d^{2}\mathbf{r}_{\perp}~\mathbf{u}$, and focus on its $x$ component, $p_{x}$, in the reference frame of the fluid (see SM for details). We measure this observable as a function of $\beta$ in Fig.~\ref{fig:momentum_swim}(b). Data are reported as purple squares. The blue triangles and the gray inverted triangles show the incompressible (vortices) and compressible (sound waves) components of $p_{x}$, respectively, obtained from the previously used decomposition of $\mathbf{u}$. Below the critical Mach number for vortex nucleation, i.e., for $\beta<\beta_{\mathrm{c}}\simeq0.35$, the net momentum $p_{x}$ is zero, as expected in this excitation-free, superfluid-flow regime. Upon exceeding $\beta_{\mathrm{c}}$, $p_{x}$ becomes negative (directed opposite to the incoming stream), which promotes the upstream propulsion of the impurity. This effect is clearly associated with the nucleation of vortices within the fluid, as $p_{x}$ is dominated by its negative incompressible contribution.

In this work, we have quantitatively investigated, both experimentally and theoretically, the dynamics of a mobile impurity in a flow of superfluid light, showing that it can swim against the superfluid current as a result of the periodic nucleation of vortex-antivortex pairs above the critical speed. Our experimental platform provides precise control over the injected flow and enables high-resolution measurements of the full optical field, allowing one to study both the impurity dynamics and the counterflow generated by the alley of vortices emitted downstream. By employing an additional imaging system to track the impurity trajectory along the medium, we directly correlate the impurity acceleration with the emission of these vortex pairs. Most importantly, by decomposing the velocity field into its compressible and incompressible components, we show that the negative impurity's momentum originates predominantly from the vortices, drawing a direct analogy with self-propulsion in active-matter systems and classical hydrodynamics. Our observations are corroborated by a theoretical model of vortex-induced propulsion in a 2D superflow. This model provides the impurity's equation of motion, establishes the condition for upstream acceleration (a requirement satisfied in our setup), and explains the physical origin of this motion: a repulsive density potential formed at the rear of the impurity. These results open new perspectives for studying the dynamics of impurities in superfluid flows, and highlight interesting connections with, e.g., self-propulsion mechanisms in active-matter systems.

\acknowledgements

We acknowledge M\'ed\'eric Argentina, Thomas Frisch, Fr\'ed\'eric H\'ebert, and Quentin Schibler for discussions. This work has been supported by Agence Nationale de la Recherche (ANR), France under Grants Nos.~ANR-21-CE47-0009 (Quantum-SOPHA), ANR-21-CE30-0008 (STLight), and ANR-24-CE47-4949 (UniQ-RingS).

\bibliography{biblio}

\end{document}


\title{Swimming against a superfluid flow: \texorpdfstring{\\}{Lg} Self-propulsion via vortex-antivortex shedding in a quantum fluid of light \texorpdfstring{\\}{Lg} Supplemental material}

\author{Myrann Baker-Rasooli}
\affiliation{Sorbonne Universit\'e, ENS Paris, Coll\`ege de France, Universit\'e PSL, CNRS, Laboratoire Kastler Brossel, F-75005 Paris, France}

\author{Tangui Aladjidi}
\affiliation{Sorbonne Universit\'e, ENS Paris, Coll\`ege de France, Universit\'e PSL, CNRS, Laboratoire Kastler Brossel, F-75005 Paris, France}

\author{Tiago D. Ferreira}
\affiliation{Universidade do Porto and INESC TEC, Centre of Applied Photonics, Rua do Campo Alegre, Porto, Portugal}

\author{Alberto Bramati}
\affiliation{Sorbonne Universit\'e, ENS Paris, Coll\`ege de France, Universit\'e PSL, CNRS, Laboratoire Kastler Brossel, F-75005 Paris, France}
\affiliation{Institut Universitaire de France (IUF)}

\author{Mathias Albert}
\affiliation{Universit\'e C\^ote d'Azur, CNRS, INPHYNI, France}
\affiliation{Institut Universitaire de France (IUF)}

\author{Pierre-\'Elie Larr\'e}
\email{pierre-elie.larre@universite-paris-saclay.fr}
\affiliation{Universit\'e Paris-Saclay, CNRS, LPTMS, 91405, Orsay, France}

\author{Quentin Glorieux}
\email{quentin.glorieux@sorbonne-universite.fr}
\affiliation{Sorbonne Universit\'e, ENS Paris, Coll\`ege de France, Universit\'e PSL, CNRS, Laboratoire Kastler Brossel, F-75005 Paris, France}
\affiliation{Institut Universitaire de France (IUF)}

\maketitle

\section{Experimental setup}

\begin{figure}[b!]
\centering
\includegraphics[width=\linewidth]{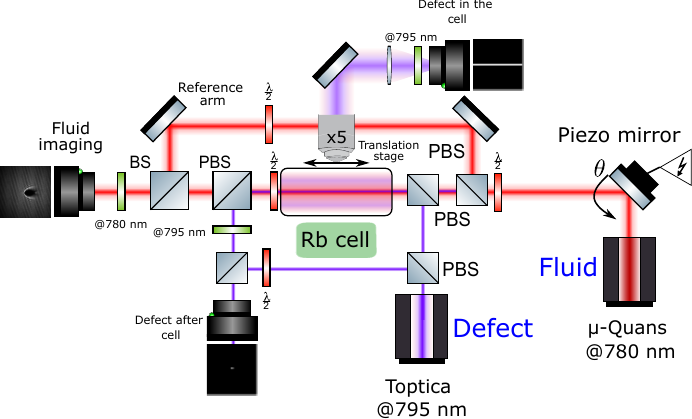}
\caption{\textbf{Experimental setup.} A 780 nm laser beam is sent into a 20 cm-long rubidium-vapor cell to generate the superfluid of light. An optical impurity is produced with a 795 nm, narrower laser beam, overlapping with the superfluid beam. It is tuned close to the D1 line with a positive detuning (\maths{\sim+2~\mathrm{GHz}}), leading to a relative linear-refractive-index change via optical pumping. The output plane of the nonlinear medium is then imaged on a camera. The relative angle between the laser beams is set by a piezo-mirror, which adjusts the fluid-beam angle at the input. A reference beam, different from the initial fluid beam, is recombined with the main beam before the camera to enable phase measurement. The side of the cell is also imaged with a microscope placed on a translation stage, and the measured fluorescence is sent to another camera.}
\label{fig:manip_detail}
\end{figure}

As shown in Fig.~\ref{fig:manip_detail},
we create a superfluid of light by propagating a 780 nm laser beam through a \maths{L=20~\mathrm{cm}}-long \textsuperscript{87}Rb-vapor cell heated to \maths{150~^{\circ}\mathrm{C}}, which acts as a nonlinear medium. The beam is slightly detuned from the rubidium D2 line (\maths{\sim-8~\mathrm{GHz}}) to have a substantial optical nonlinearity. An optical impurity is introduced by overlapping an auxiliary beam with the beam creating the superfluid, along the \maths{z} axis at the center of the cell. This impurity beam has a waist \maths{\sigma\simeq7.5\xi} of the order of a few healing lengths \maths{\xi\simeq50~\mu\mathrm{m}}, much smaller than the overall transverse size \maths{\sim100\xi} of the superfluid. To make the impurity act as a transverse potential barrier for the superfluid, the 795 nm impurity beam is tuned close to the D1 line with a positive detuning (\maths{\sim+2~\mathrm{GHz}}), leading to a relative linear-refractive-index change via optical pumping. The transverse-flow velocity past this potential barrier is optically controlled by only adjusting the superfluid-beam angle \maths{\mathbf{k}_{\perp}} with a piezo-mirror, while the impurity beam is on-axis.

Using a dual-imaging system, we simultaneously measure the fluorescence of the beams from the side and record the amplitude and phase of the superfluid of light at the exit of the rubidium cell. These are recorded using an off-axis interferometer. The side of the cell is imaged with a microscope located on a translation stage, and the measured fluorescence is sent to an other camera, making it possible to reconstruct, as a function of the propagation time \maths{z}, the impurity-beam trajectory.

\section{Superfluid velocity}
\begin{figure}[b!]
\centering
\includegraphics[width=\linewidth]{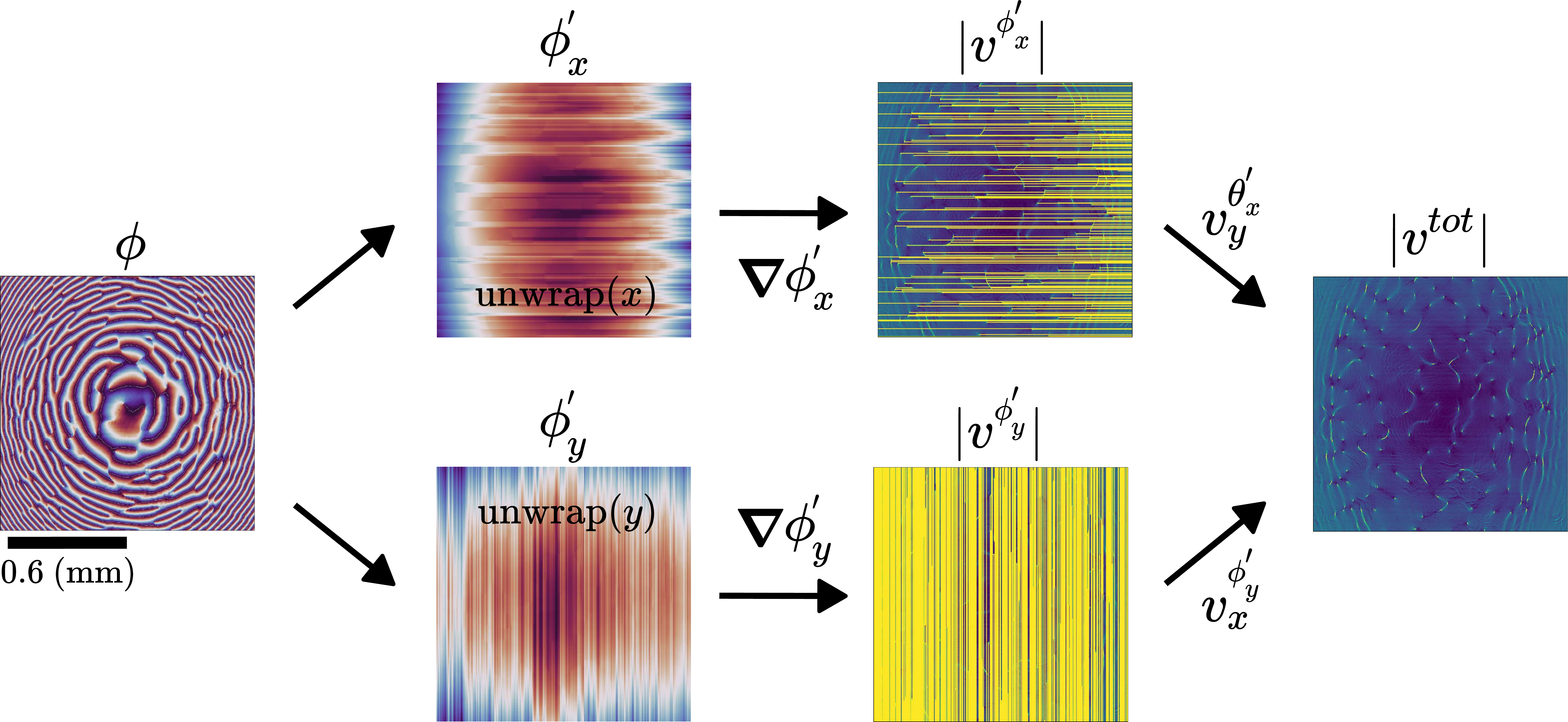}
\caption{\textbf{From the phase to the velocity of the superfluid.}
The total velocity, reconstructed from the combination of each gradient component calculated using the 1D unwrapped-phase method.}
\label{fig:velo_tot}
\end{figure}
From the superfluid-phase map \maths{\phi=\phi(\mathbf{r}_{\perp})}, we reconstruct the total velocity field of the flow, \maths{\mathbf{v}(\mathbf{r}_{\perp})=\nabla_{\perp}\phi(\mathbf{r}_{\perp})/k_{0}}.
To compute the velocity map, we must first perform a 2D unwrapping of the phase.
However, because we work with discretized data arrays, phase singularities (i.e., $2\pi$ windings) create numerical discontinuities that prevent an accurate reconstruction \cite{glorieux2025}.
To circumvent this issue, the phase is unwrapped separately along each axis, yielding \maths{\phi_{x}'} and \maths{\phi_{y}'}.
The total velocity field is then obtained by combining the gradient components computed from these two independently unwrapped phase profiles, as illustrated in Fig.~\ref{fig:velo_tot}.

\section{Sound velocity}

When entering the rubidium cell, the fluid beam undergoes a sudden increase in energy as a function of the propagation time \maths{z}, due to the presence of the impurity potential. This energy is typically redistributed into counterpropagating sound waves with velocities \maths{\pm c_{\mathrm{s}}} in the transverse \maths{\mathbf{r}_{\perp}}-plane, a process analogous to quench dynamics in weakly interacting quantum gases. We measure the speed of sound \maths{c_{\mathrm{s}}} in the superfluid by analyzing the transverse separation of these waves at final time \maths{z=L}. Specifically, half their transverse separation, \maths{r_{\mathrm{s}}}, is related to \maths{c_{\mathrm{s}}} by the simple relation \maths{r_{\mathrm{s}}=c_{\mathrm{s}}L}.

\begin{figure}[b!]
\centering
\includegraphics[width=\columnwidth]{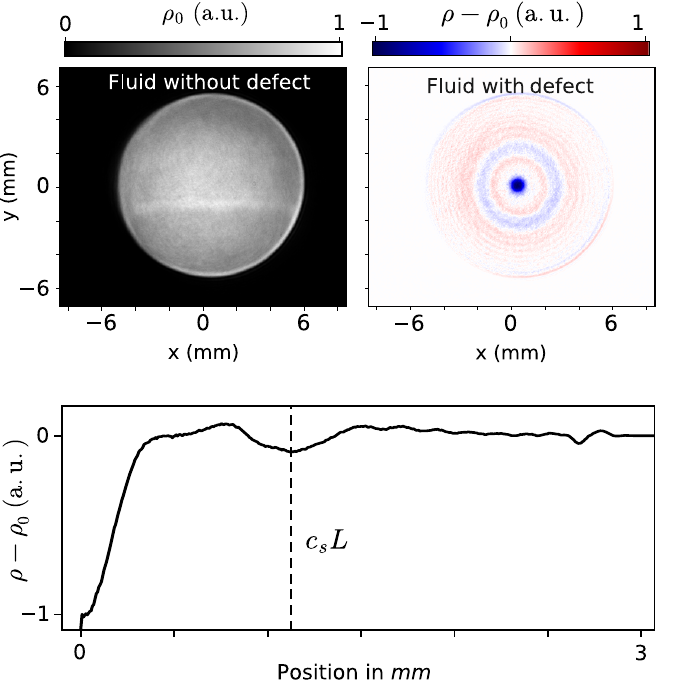}
\caption{\textbf{Measuring the speed of sound.} Top: The fluid density without the impurity, \maths{\rho_{0}}, is subtracted from the fluid density with the impurity, \maths{\rho}, revealing the formation of an acoustic perturbation (blue annulus) expanding with velocity \maths{c_{\mathrm{s}}} in the transverse plane. Bottom: This speed of sound is deduced from the location \maths{r_{\mathrm{s}}=c_{\mathrm{s}}L} of the acoustic dip (vertical dotted line) in the averaged \maths{\rho-\rho_{0}} at final propagation time \maths{z=L}.}
\label{fig:cs_quench}
\end{figure}

This measurement is performed in the density difference \maths{\rho-\rho_{0}}, where \maths{\rho} is the superfluid density with the impurity, and \maths{\rho_{0}} is the density without it. Typical images are shown in Fig.~\ref{fig:cs_quench}
(top panels). We then perform an azimuthal average of this density difference, centered on the impurity, which is shown as a function of the radial position in the bottom panel. From the position \maths{r_{\mathrm{s}}} of the sound wave (corresponding to the density dip and indicated by the red vertical line), we extract the speed of sound \maths{c_{\mathrm{s}}} from \maths{r_{\mathrm{s}}=c_{\mathrm{s}}L}. Since this measurement is performed at the exit of the hot vapor and the camera collects light cumulatively, the procedure does not provide the instantaneous speed of sound but rather its value averaged over the entire propagation, taking into account light absorption through the medium.

\section{Vortex-shedding frequency}
\begin{figure}[h!]
\centering
\includegraphics[width=0.85\columnwidth]{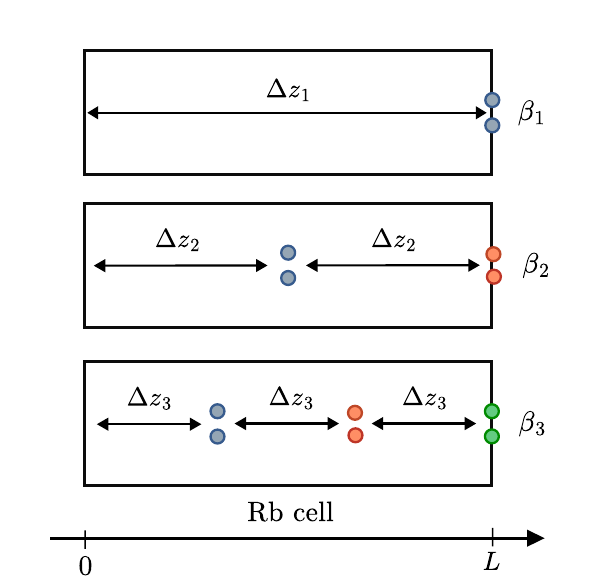}
\caption{\textbf{Periodic vortex shedding along the cell.} For each emission event \maths{i\geqslant1} at the output of the cell, the shedding frequency \maths{1/\Delta z} is inferred by dividing the Rb cell into \maths{i} segments. The first (second, third) pair at \maths{\beta=\beta_{1}=\beta_{\mathrm{c}}} (\maths{\beta_{2}}, \maths{\beta_{3}}) is in blue (red, green).}
\label{fig:tuto_freqency}
\end{figure}
We assume that the shedding of vortex-antivortex pairs occurs periodically and consider that each new pair detected by the camera is produced at the exit of the cell, supposing it forms instantaneously. Figure~\ref{fig:tuto_freqency}
provides a schematic view of the vortex pairs periodically emitted along the rubidium cell as the Mach number of the flow, \maths{\beta}, is increased from the critical value \maths{\beta_{1}=\beta_{\mathrm{c}}} required for nucleation of the first pair. Each time the camera detects the \maths{i}th pair at \maths{\beta=\beta_{i}} (\maths{i\geqslant1}), the time elapsed since the observation of the \maths{(i-1)}th pair at \maths{\beta=\beta_{i-1}} is necessarily \maths{\Delta z_{i}=L/i}. From the data \maths{\{\beta_{i},\Delta z_{i}\}}, we infer the vortex-shedding frequency \maths{1/\Delta z} as a function of the flow Mach number \maths{\beta}, which exhibits a linear trend very consistent with previous cold-atom experiments~\cite{kwon2015periodic} [see Fig.~2(c) of the main text].

\section{Counterflow impurity motion: Theory}

\begin{figure}[t!]
\centering
\includegraphics[scale=1]{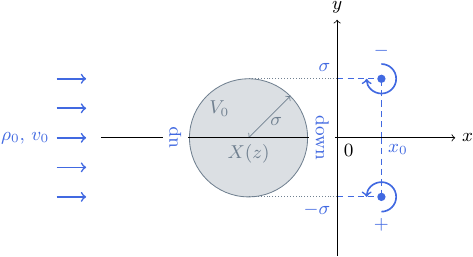}
\caption{\textbf{Theoretical picture of the system.} A superflow (blue) impinges upon a mobile impurity (gray), which self-propels upstream by vortex-antivortex shedding downstream.}
\label{Fig:System}
\end{figure}

The studied system is composed of two components: a 2D (\maths{\mathbf{r}_{\perp}=x\hat{\mathbf{x}}+y\hat{\mathbf{y}}}) superfluid of light flowing in the positive-\maths{x} direction with incoming density \maths{\rho_{0}} and velocity \maths{v_{0}}; and an optical impurity embedded within the vapor, modeled as a repulsive, localized potential \maths{V(\mathbf{r}_{\perp}-X\hat{\mathbf{x}})} with maximum amplitude \maths{V_{0}}, characteristic radius \maths{\sigma}, and time-dependent center-of-mass position \maths{X(z)} along the incoming stream. This is illustrated in Fig.~\ref{Fig:System},
which also schematically shows one of the vortex-antivortex pairs responsible for propelling the impurity upstream. The notations are the ones used in the main text.

The superfluid is described by an electric field with a slowly varying envelope \maths{E(\mathbf{r}_{\perp},z)} obeying the NLS equation [Eq.~(1) in the main text]. Its Hamiltonian, in the absence of the impurity, is given by
\begin{equation}
\label{Eq:Hfluid}
H_{\mathrm{f}}=\alpha\int d^{2}\mathbf{r}_{\perp}~\bigg(\frac{|\nabla_{\perp}E|^{2}}{2k_{0}}-\frac{k_{0}n_{2}\rho^{2}}{2}\bigg),
\end{equation}
where the first, mass term originates from paraxial diffraction (\maths{k_{0}} is the propagation constant of the laser beam creating the superfluid), and the last, interaction term is due to the Kerr nonlinearity of the vapor (\maths{n_{2}\rho} is the Kerr shift in the local refractive index of the vapor, with \maths{n_{2}<0} and \maths{\rho=|E|^{2}} the superfluid density). The constant \maths{\alpha} ensures that \maths{H_{\mathrm{f}}} has the correct units of a wave number in the language of superfluids of light [see discussion after Eq.~\eqref{Eq:Xz}]. Here, we focus solely on the motion of the impurity's center of mass, disregarding any temporal change to its shape. Therefore, the impurity can be modeled as a point particle whose dynamics, in the absence of the superfluid, is obtained from the kinetic Hamiltonian
\begin{equation}
\label{Eq:Himpurity}
H_{\mathrm{i}}=\frac{P^{2}}{2k_{\mathrm{i}}}.
\end{equation}
In this equation, \maths{P(z)} denotes the momentum of the impurity's center of mass, and \maths{k_{\mathrm{i}}} is its mass. Finally, the fluid and the impurity interact via the standard Hamiltonian
\begin{equation}
\label{Eq:Hint}
H_{\mathrm{f,i}}=\alpha\int d^{2}\mathbf{r}_{\perp}~V(\mathbf{r}_{\perp}-X\hat{\mathbf{x}})\rho,
\end{equation}
where \maths{\alpha} is the same normalization constant as in Eq.~\eqref{Eq:Hfluid}. The equations of motion obtained from the full Hamiltonian, \maths{H=H_{\mathrm{f}}+H_{\mathrm{i}}+H_{\mathrm{f,i}}}, are \maths{\partial_{z}(\sqrt{\alpha}E)=\delta H/\delta(i\sqrt{\alpha}E^{\ast})} (and its complex conjugate), \maths{\dot{X}\equiv dX/dz=\partial_{P}H}, and \maths{\dot{P}=-\partial_{X}H}, which leads to the coupled dynamical equations
\begin{align}
\label{Eq:NLS}
i\partial_{z}E&=\bigg[{-}\frac{\nabla_{\perp}^{2}}{2k_{0}}+V(\mathbf{r}_{\perp}-X\hat{\mathbf{x}})-k_{0}n_{2}\rho\bigg]E, \\
\label{Eq:Newton}
k_{0}\ddot{X}&=F=\alpha\int d^{2}\mathbf{r}_{\perp}~\partial_{x}V(\mathbf{r}_{\perp}-X\hat{\mathbf{x}})\rho.
\end{align}
Equation~\eqref{Eq:NLS} is nothing but Eq.~(1) of the main text, and Eq.~\eqref{Eq:Newton} is Newton's second law of motion for the impurity's center of mass, where the force \maths{F} coincides with the drag force exerted by the superfluid on the impurity (see, e.g., Ref.~\cite{Pavloff2002}). This final equation for the impurity's trajectory along the incoming stream, \maths{X=X(z)}, is highly nonlinear but can be approximately solved under certain simplifying physical hypotheses, as detailed below. In the following, we analyze a single cycle of vortex-antivortex shedding. The full trajectory is then obtained by using the final center-of-mass position and velocity in each cycle as the initial conditions for the next.

The core object in Eq.~\eqref{Eq:Newton} is the drag force experienced by the impurity, \maths{F}, which we evaluate in the following way:
\begin{align}
\label{Eq:F-a}
F&=-\alpha\int d^{2}\mathbf{r}_{\perp}~V(\mathbf{r}_{\perp})\partial_{x}\rho(\mathbf{r}_{\perp}+X\hat{\mathbf{x}},z) \\
\label{Eq:F-b}
&\approx-\alpha\ell V_{0}(\rho_{\mathrm{down}}-\rho_{\mathrm{up}}).
\end{align}
Equation~\eqref{Eq:F-a} is obtained from a straightforward change of variables and integration by parts. In approximation~\eqref{Eq:F-b}, the force is estimated by assuming that the typical radius \maths{\sigma} of the impurity is large enough to dominate the geometry of its cross-section along the \maths{y} axis, and by considering that its potential, in the reference frame of its center of mass, is equal to \maths{V_{0}} for \maths{|x|<\sigma} and zero otherwise. This simplifies the integral over \maths{y} to a multiplication by the typical transverse width \maths{\ell} of the incoming flux, and transforms \maths{V\partial_{x}\rho} into \maths{V_{0}} multiplied by the difference in density between the downstream (\maths{x=X+\sigma}) and upstream (\maths{x=X-\sigma}) poles of the impurity. Then, to evaluate \maths{\rho_{\mathrm{up}}} and \maths{\rho_{\mathrm{down}}}, we use the Madelung equation for the superfluid's velocity field \maths{\mathbf{v}=\nabla_{\perp}\mathrm{arg}(E)/k_{0}}, which is the real part of the NLS equation~\eqref{Eq:NLS} (see, e.g., Ref.~\cite{Pitaevskii2016}). The dispersionless limit for this equation is appropriate in the configuration of our experiment, where \maths{\sigma} is larger than the healing length \maths{\xi=1/(2k_{0}^{2}|n_{2}|\rho_{0}^{\vphantom{2}})^{1/2}}. In addition, we assume that the velocity field surrounding the impurity does not vary significantly over one cycle of vortex emission. Therefore, the densities \maths{\rho_{\mathrm{up}}} and \maths{\rho_{\mathrm{down}}} can be evaluated in terms of the velocities \maths{\mathbf{v}_{\mathrm{up}}} and \maths{\mathbf{v}_{\mathrm{down}}}, respectively, from the Bernoulli-type law
\begin{equation}
\label{Eq:Bernouche}
\frac{k_{0}^{\vphantom{2}}\mathbf{v}_{\mathrm{up},\mathrm{down}}^{2}}{2}-k_{0}n_{2}\rho_{\mathrm{up},\mathrm{down}}=\frac{k_{0}^{\vphantom{2}}v_{0}^{2}}{2}-k_{0}n_{2}\rho_{0}.
\end{equation}
Upstream, we roughly approximate the velocity of the flow to its asymptotic value:
\begin{equation}
\label{Eq:vup}
\mathbf{v}_{\mathrm{up}}\approx v_{0}\hat{\mathbf{x}}.
\end{equation}
Downstream, the emitted vortex-antivortex pair superimposes on the background flow, which results in a total velocity \maths{\mathbf{v}_{\mathrm{down}}\approx v_{0}\hat{\mathbf{x}}+\mathbf{v}_{\mathrm{v}}}. A point-vortex model, which can be obtained from the 2D Hamiltonian~\eqref{Eq:Hfluid}~\cite{Skipp2023}, makes it possible to evaluate \maths{\mathbf{v}_{\mathrm{v}}} in the form (see, e.g., Ref.~\cite{Dalibard2017})
\begin{equation}
\label{Eq:vvav}
\mathbf{v}_{\mathrm{v}}=\sum_{\nu=\pm1}\frac{\Gamma_{\nu}}{2\pi}\bigg({-}\frac{y-y_{\nu}}{|\mathbf{r}_{\perp}-\mathbf{r}_{\nu}|^{2}}\hat{\mathbf{x}}+\frac{x-x_{\nu}}{|\mathbf{r}_{\perp}-\mathbf{r}_{\nu}|^{2}}\hat{\mathbf{y}}\bigg),
\end{equation}
where \maths{\Gamma_{\nu}=2\pi\nu/k_{0}} is the circulation of the flow velocity around the vortex with charge \maths{\nu=\pm1} and position \maths{\mathbf{r}_{\nu}=x_{\nu}\hat{\mathbf{x}}+y_{\nu}\hat{\mathbf{y}}}, where \maths{x_{\nu}=x_{-\nu}=x_{0}} and \maths{y_{\nu}=-\nu\sigma} (see Fig.~\ref{Fig:System}). This leads to, at the center of the pair for simplicity (\maths{x=x_{0}} and \maths{y=0}),
\begin{equation}
\label{Eq:vdown}
\mathbf{v}_{\mathrm{down}}\approx\bigg(v_{0}-\frac{2}{k_{0}\sigma}\bigg)\hat{\mathbf{x}}.
\end{equation}
Insertion of~\eqref{Eq:vup} and~\eqref{Eq:vdown} into Eq.~\eqref{Eq:Bernouche} yields
\begin{align}
\label{Eq:rhoup}
\rho_{\mathrm{up}}&\approx\rho_{0}, \\
\label{Eq:rhodown}
\rho_{\mathrm{down}}&\approx\rho_{0}+\Delta\rho,\quad\Delta\rho=2\sqrt{2}\rho_{0}\frac{\xi}{\sigma}\bigg(\beta-\sqrt{2}\frac{\xi}{\sigma}\bigg),
\end{align}
where \maths{\beta=v_{0}/c_{\mathrm{s}}} is the Mach number of the incoming flow, \maths{c_{\mathrm{s}}=(|n_{2}|\rho_{0})^{1/2}} being the speed of sound far from the impurity.

When the expressions for the upstream and downstream densities from Eqs.~\eqref{Eq:rhoup} and~\eqref{Eq:rhodown} are substituted into the drag force~\eqref{Eq:F-b}, Newton's law~\eqref{Eq:Newton} for the motion of the impurity's center of mass along the incoming stream integrates easily to give
\begin{equation}
\label{Eq:Xz}
X(z)\approx Az^{2}+\dot{X}_{0}z+X_{0},\quad A=-\frac{\alpha\ell V_{0}\Delta\rho}{k_{\mathrm{i}}}.
\end{equation}
In these equations, \maths{X_{0}} and \maths{\dot{X}_{0}} are the initial position and velocity, respectively, of the impurity for the cycle in question (for the very first cycle, both \maths{X_{0}} and \maths{\dot{X}_{0}} are zero). The acceleration \maths{A}, given by the second of Eqs.~\eqref{Eq:Xz}, depends on several key quantities, starting with the normalization constant \maths{\alpha} introduced in Eq.~\eqref{Eq:Hfluid}. This constant incorporates the microscopic parameters of the laser and medium used to produce the superfluid of light. Because we do not derive its exact form here, we treat it as a fitting parameter to match the parabolic model~\eqref{Eq:Xz} to the measured trajectory (see details in the main text). Given that \maths{\alpha\simeq3~\mathrm{V}^{-2}} is found to be of a nonextreme order of magnitude, we are confident that the remaining dependencies in the formula for \maths{A} are sufficient to explain its underlying physics. These remaining factors include the transverse size \maths{\ell\simeq19\xi} of the incoming flux (\maths{\sim} impurity diameter plus two double vortex radii), the impurity's amplitude \maths{V_{0}\simeq7.9~\mathrm{m}^{-1}},
and, most importantly, the downstream density variation \maths{\Delta\rho}, whose expression in terms of the impurity's radius \maths{\sigma} and the Mach number \maths{\beta} is provided in~\eqref{Eq:rhodown}. From the latter dependence, we deduce that the acceleration \maths{A} is negative---i.e., the impurity is propelled upstream---if
\begin{equation}
\label{Eq:Counterflow}
\beta>\sqrt{2}\frac{\xi}{\sigma}.
\end{equation}
This condition is met at a sufficiently high flow velocity compared to the speed of sound, or for a sufficiently wide impurity compared to the healing length. In our experiment, \maths{\sqrt{2}\xi/\sigma\simeq0.19} (\maths{\sigma\simeq7.5\xi}), which is below the critical Mach number for vortex nucleation that we measure, \maths{\beta_{\mathrm{c}}\simeq0.35}. Therefore, constraint~\eqref{Eq:Counterflow} is satisfied in our setup (\maths{\beta\gtrsim\beta_{\mathrm{c}}}), quantitatively conditioning upstream motion of the impurity. Interestingly, this is associated with the formation of a density hump downstream, \maths{\Delta\rho>0}, most certainly due to the clockwise (for \maths{y>0}) and anticlockwise (\maths{y<0}) vorticity windings (see Fig.~\ref{Fig:System}), which together drag light intensity to the rear of the impurity. This density hump acts as a repulsive potential for the impurity, which then moves upstream to minimize its energy.
\begin{figure}[b!]
\centering
\includegraphics[width=0.85\linewidth]{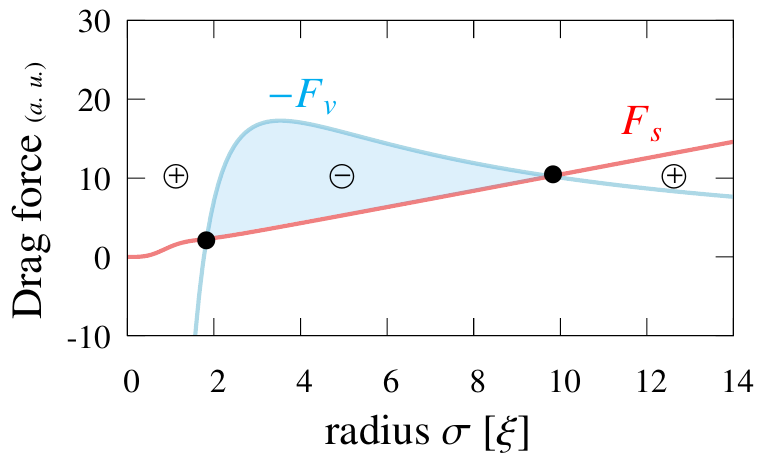}
\caption{\textbf{Sound and vortex contributions to the total drag.} The force \maths{F_\mathrm{s}} is the contribution from the sound waves emitted upstream for \maths{\beta\gtrsim1} [Eq.~\eqref{Eq:Fs}], and \maths{F_\mathrm{v}} is the one from the vortex pairs emitted downstream [Eqs.~\eqref{Eq:F-b},~\eqref{Eq:rhoup}, and~\eqref{Eq:rhodown}]. Both forces are plotted in arbitrary units against the impurity's radius \maths{\sigma} in units of the healing length. The shaded area corresponds to the regime where self-propulsion against the flow is possible at sonic velocities, characterized by a total drag force \maths{F_{\mathrm{v}}+F_{\mathrm{s}}<0}.}
\label{Fig:Diag_phase}
\end{figure}
Counterflow motion of the impurity is explained by the shedding of vortices downstream, once the Mach number \maths{\beta} exceeds the critical value \maths{\beta_{\mathrm{c}}\simeq0.35}. When \maths{\beta} is further increased, other excitations are stimulated in the superfluid, such as \textit{sound waves} upstream. As evidenced in the main text, this occurs for \maths{\beta} approaching unity, surely for \maths{\beta\gtrsim1}, i.e., when the flow velocity \maths{v_{0}} exceeds the speed of sound \maths{c_{\mathrm{s}}}. In this velocity regime, the work imparted by this sonic radiation manifests as a positive drag force \maths{F_{\mathrm{s}}}, opposite to the negative vortex force \maths{F\equiv F_{\mathrm{v}}} estimated in Eqs.~\eqref{Eq:F-b},~\eqref{Eq:rhoup}, and~\eqref{Eq:rhodown}. For a Gaussian potential \maths{V(\mathbf{r}_{\perp}^{\vphantom{2}})=V_{0}\exp[-\mathbf{r}_{\perp}^{2}/(2\sigma^{2})]}, the dependence of \maths{F_{\mathrm{s}}} on \maths{\beta} and \maths{\sigma} can be easily estimated provided \maths{V_{0}} is smaller than the inverse nonlinear length, \maths{1/z_{\mathrm{NL}}}. We find~\cite{Ferreira2022}
\begin{equation}
\label{Eq:Fs}
F_{\mathrm{s}}\propto\frac{\bar{\beta}^{2}}{\beta}\frac{\sigma^{4}}{\xi^{4}}e^{-\bar{\beta}^{2}\sigma^{2}/\xi^{2}}\Big[I_{0}(\bar{\beta}^{2}\sigma^{2}/\xi^{2})-I_{1}(\bar{\beta}^{2}\sigma^{2}/\xi^{2})\Big],
\end{equation}
where \maths{\bar{\beta}=[2(\beta^{2}-1)]^{1/2}} and \maths{I_{n}} is the modified Bessel function of the first kind. Assuming vortices still contribute to the drag at these sonic velocities and for such an impurity, motion against the incoming stream is only possible when the sound contribution \maths{F_{\mathrm{s}}} is not large enough to overcome the vortex contribution \maths{F_{\mathrm{v}}}, specifically when \maths{F_{\mathrm{s}}<-F_{\mathrm{v}}} (see Fig.~\ref{Fig:Diag_phase}).
This situation contrasts with the experiment by Michel \textit{et al.}~\cite{Michel2018}, where the impurity strength \maths{V_{0}} was too small compared to \maths{1/z_{\mathrm{NL}}} to generate vortices, resulting in a single positive force \maths{F_{\mathrm{s}}} due to upstream radiation of sound waves, and then downstream motion of the mobile impurity.

\section{Helmholtz decomposition}

In the main text, we analyze separately the ``comp''ressible (divergent) and ``inc''ompressible (rotational) components of the density-weighted velocity field, defined as
\begin{equation}
\mathbf{u}(\mathbf{r}_{\perp})=\sqrt{\rho(\mathbf{r}_{\perp})}\mathbf{v}(\mathbf{r}_{\perp}).
\end{equation}
These two contributions appear in the following Helmholtz decomposition:
\begin{align}
\mathbf{u}(\mathbf{r}_{\perp})&=\mathbf{u}^{\mathrm{comp}}(\mathbf{r}_{\perp})+\mathbf{u}^{\mathrm{inc}}(\mathbf{r}_{\perp}), \\
\mathbf{u}^{\mathrm{comp}}(\mathbf{r}_{\perp})&=\nabla_{\perp}\Phi(\mathbf{r}_{\perp}), \\
\mathbf{u}^{\mathrm{inc}}(\mathbf{r}_{\perp})&=\nabla_{\perp}\times\mathbf{A}(\mathbf{r}_{\perp}),
\end{align}
where \maths{\Phi} and \maths{\mathbf{A}} are scalar and vector potentials, respectively. In practice, the incompressible velocity is obtained by subtracting the compressible component from the total density-weighted velocity, \maths{\mathbf{u}^{\mathrm{inc}}=\mathbf{u}-\mathbf{u}^{\mathrm{comp}}}, with \maths{\mathbf{u}^{\mathrm{comp}}} obtained from its Fourier components:
\begin{align}
\mathbf{u}^{\mathrm{comp}}(\mathbf{r}_{\perp})&=\mathrm{FT}^{-1}[i\mathbf{k}_{\perp}U_{\Phi}(\mathbf{k}_{\perp})](\mathbf{r}_{\perp}), \\
U_{\Phi}(\mathbf{k}_{\perp})&=-i\frac{\mathbf{k}_{\perp}}{k_{\perp}^{2}}\cdot\mathrm{FT}[\mathbf{u}(\mathbf{r}_{\perp})](\mathbf{k}_{\perp}).
\end{align}


\section{Net momentum of the flow}

To determine the contribution of vortices to the net momentum of the fluid, we make use of the Helmholtz decomposition described above, which separates the incompressible component, associated with vortices, from the compressible component, associated with sound waves, of the density-weighted velocity \maths{\mathbf{u}}. Figure~\ref{fig:momentum}(a)
\begin{figure}[t!]
\centering
\includegraphics[width=\columnwidth]{net_momentum.pdf}
\caption{\textbf{Contributions to the net momentum of the fluid.} \textbf{a} -- Norm of the density-weighted velocity map at \maths{\beta=1.1} for the incompressible (top) and compressible (bottom) components, shown in the laboratory (left) and fluid (right) reference frames, respectively.
The associated streamlines are displayed on each image.
\textbf{b} -- Final impurity position measured at the exit of the cell for different values of the Mach number \maths{\beta}.
\textbf{c} -- Total net momentum against $\beta$ in the laboratory frame, with ($p$) and without ($p_0$) the impurity, shown by the green and gray dots, respectively.
\textbf{d} -- Fluid-frame net momentum for the total (green), incompressible (orange), and compressible (blue) components of the density-weighted velocity.}
\label{fig:momentum}
\end{figure}
shows the norm of the incompressible (top panels) and compressible (bottom) contributions for \maths{\beta=1.1}. As explained above, the incompressible field is obtained by subtracting the compressible contribution from the total density-weighted velocity. The left and right panels display the velocity fields in the laboratory and fluid frames, respectively. The change of reference frame is performed by subtracting the phase of the fluid in the absence of the impurity, using the same value of \maths{\beta}. In the fluid frame, we observe that the vortex contribution is predominantly directed towards the negative-\maths{x} direction, in contrast to the compressible part, which exhibits contributions in all directions.

To quantitatively show that this behavior is directly related to the upstream displacement of the impurity, as observed in Fig.~\ref{fig:momentum}(b), we measure the total net momentum along the \maths{x} axis, defined as
\begin{equation}
\label{Eq:p}
p = \int d^{2}\mathbf{r}_{\perp}~u_{x}(\mathbf{r}_{\perp}),\quad u_{x}(\mathbf{r}_{\perp})=\sqrt{\rho(\mathbf{r}_{\perp})}v_{x}(\mathbf{r}_{\perp}).
\end{equation}
We denote by $p_{0}$ the net momentum of the fluid in the absence of the impurity, and by $p$ its counterpart in the presence of the impurity. 
It is worth noting that we use~\eqref{Eq:p} as a proxy, the actual, physical momentum being rigorously defined as the integral of the current density \maths{\rho v_{x}}.

In the laboratory frame, and without the impurity, we observe a linear increase of the net momentum, as shown in Fig.~\ref{fig:momentum}(c) by the gray dots.
This behavior changes in the presence of the impurity: a clear bifurcation appears once \maths{\beta=\beta_\mathrm{c} \simeq 0.35}, i.e., when the first vortex-antivortex pair is emitted.
By subtracting $p_{0}$ from $p$, we place ourselves in the reference frame of the fluid and separate the vortex and acoustic contributions to the total momentum.
The different components of the momentum are shown in Fig.~\ref{fig:momentum}(d).
The green squares, orange triangles, and blue triangles correspond to the total, incompressible, and compressible contributions, respectively.
These values show a clear negative trend, with the dominant contribution to the total momentum coming from the vortices, while the compressible component remains almost negligible.
For \maths{\beta > 0.8}, as discussed in the main text, additional phenomena appear, such as radiation of sound waves ahead of the impurity.
At this point, the balance between the compressible and incompressible components starts to change, since vortex emission has ceased.

\clearpage